# Ideal *n*-body correlations with massive particles

R.G. Dall[(1)], A.G. Manning[(1)], S.S. Hodgman[(1)], Wu RuGway[(1)], K.V. Kheruntsyan[(2)], and A.G. Truscott [(1)*]

[(1)]*Research School of Physics and Engineering, Australian National University, Canberra, ACT 0200, Australia*

[(2)]*The University of Queensland, School of Mathematics and Physics, Brisbane, QLD 4072, Australia*

*e-mail: andrew.truscott@rsphysse.anu.edu.au

**In 1963 Glauber introduced the modern theory of quantum coherence [1], which extended the concept of first-order (one-body) correlations, describing phase coherence of classical waves, to include higher-order (*n*-body) quantum correlations characterizing the interference of multiple particles. Whereas the quantum coherence of photons is a mature cornerstone of quantum optics, the quantum coherence properties of massive particles remain largely unexplored. To investigate these properties, here we use a uniquely correlated [2] source of atoms that allows us to observe *n*-body correlations up to the sixth-order at the ideal theoretical limit (*n*!). Our measurements constitute a direct demonstration of the validity of one of the most widely used theorems in quantum many-body theory—Wisck's theorem [3]—for a thermal ensemble of massive particles. Measurements involving *n*-body correlations may play an important role in the understanding of thermalization of isolated quantum systems [4] and the thermodynamics of exotic many-body systems, such as Efimov trimers [5].**

Glauber's modern theory of optical coherence and the famous Hanbury Brown—Twiss effect [6] were pivotal in the establishment of the field of quantum optics. Importantly, the definition of a coherent state required coherence to all orders, which for example distinguishes a monochromatic but incoherent thermal source of light from a truly coherent source such as a laser. Higher-order correlation functions therefore provide a more rigorous test of coherence.

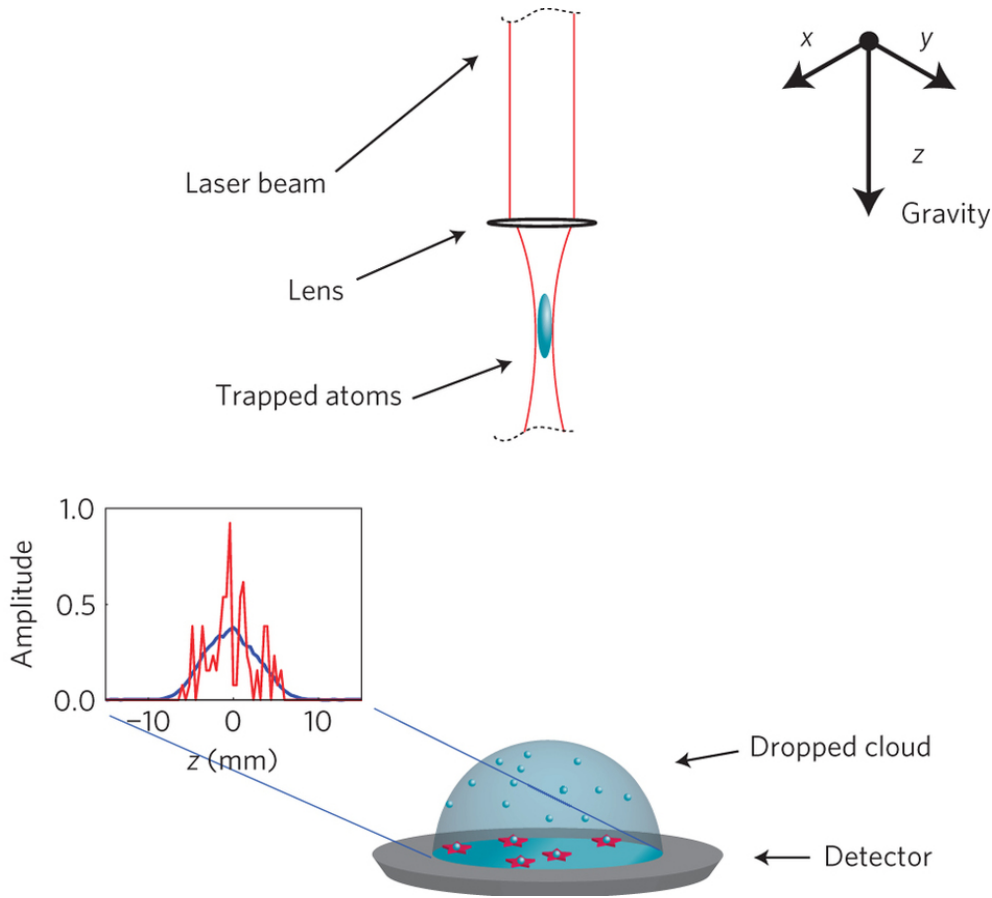

**Figure 1. Schematic of the experiment.** An ultracold cloud of atoms is confined in a tightly focused laser beam. When released the cloud drops ~85 cm where its correlation properties are measured via a single atom detector. Shown in the inset are the mean (dashed blue line) and a typical single (solid red line) longitudinal profile obtained in the experiment.

Higher-order correlations, characterised by an *n*-body correlation function $g^{(n)}$, are of general interest and have been investigated in many fields of physics including astronomy [6], particle physics [7], quantum optics [8], and quantum atom optics [9]. In particular they have been a fruitful area of research in the field of quantum optics where they have been used to investigate the properties of laser light, including heralded single photons [10], and the statistics of parametric down-conversion sources [11]. State-of-the-art quantum optics experiments have measured photon correlation functions up to sixth order for quasi-thermal sources [8], allowing the possibility of performing full quantum state tomography [12].

Higher-order correlations experiments with massive particles are currently approaching the same level of maturity as with photons. To date, experiments have directly observed correlations up to fourth order with single atom sensitive detection techniques for ultracold atomic bosons [9,13-14], and second-order correlations for an atomic source of fermions [15] demonstrating the uniquely quantum mechanical property of atom-atom antibunching. Alternative, indirect techniques have also been employed to investigate higher-order correlations, including the measurements of two-body (photoassociation [16]) and three-body [17] loss rates that are sensitive, respectively, to second- and third-order correlation functions. Interestingly, fermionic atom pairs [18] and fermionic antibunching [19] have also been observed in the atomic shot noise of absorption images.

In 1963, Glauber predicted that the maximal value of the same-point normalised *n*-body correlation function $g^{(n)}$ for thermal light is directly related to the order of the function by a simple relationship *n*! [1]. This *n*! dependence is a consequence of Wick's theorem [3], which enables higher-order correlations to be expressed using products of one-body correlation functions. The applicability of Wick's theorem is not limited to just correlation functions for light, it has been also applied to many fields; for example, it is commonly used in radio-astronomy, nuclear physics [7], and generally in quantum field theory [3]. The validity of Wick's theorem has been demonstrated with thermal photons, however, to date there has been no direct measurements demonstrating its validity to higher orders for massive particles.

The (unnormalised) two-body spatial correlation function $G^{(2)}(\mathbf{r}_1,\mathbf{r}_2) = \langle \hat{\Psi}^+(\mathbf{r}_1)\hat{\Psi}^+(\mathbf{r}_2)\hat{\Psi}(\mathbf{r}_2)\hat{\Psi}(\mathbf{r}_1) \rangle$ can be expressed as the probability of detecting two particles simultaneously at two particular locations $\mathbf{r}_1$ and $\mathbf{r}_2$, where $\hat{\Psi}^+(\mathbf{r})$ and $\hat{\Psi}(\mathbf{r})$ are the field creation and annihilation operators. When normalised by the product of atomic densities $\rho(\mathbf{r}_i) = \langle \hat{\Psi}^+(\mathbf{r}_i)\hat{\Psi}(\mathbf{r}_i) \rangle$ at respective locations, the two-body correlation function for a thermal source can be expressed mathematically in terms of the one-body correlation function $G^{(1)}(\mathbf{r}_1,\mathbf{r}_2) = \langle \hat{\Psi}^+(\mathbf{r}_1)\hat{\Psi}(\mathbf{r}_2) \rangle$:

$$g^{(2)}(\mathbf{r}_1,\mathbf{r}_2) = 1 + \frac{|G^{(1)}(\mathbf{r}_1,\mathbf{r}_2)|^2}{\rho(\mathbf{r}_1)\rho(\mathbf{r}_2)}.$$

Likewise, the three-body correlation function can be written as a nested series, allowing it to be rewritten in terms of the lower-order correlation functions (Wick's theorem):

$$g^{(3)}(\mathbf{r}_1,\mathbf{r}_2,\mathbf{r}_3) = 1 + \frac{|G^{(1)}(\mathbf{r}_1,\mathbf{r}_2)|^2}{\rho(\mathbf{r}_1)\rho(\mathbf{r}_2)} + \frac{|G^{(1)}(\mathbf{r}_2,\mathbf{r}_3)|^2}{\rho(\mathbf{r}_2)\rho(\mathbf{r}_3)} + \frac{|G^{(1)}(\mathbf{r}_3,\mathbf{r}_1)|^2}{\rho(\mathbf{r}_3)\rho(\mathbf{r}_1)} + 2\,\mathrm{Re}\,\frac{G^{(1)}(\mathbf{r}_1,\mathbf{r}_2)G^{(1)}(\mathbf{r}_2,\mathbf{r}_3)G^{(1)}(\mathbf{r}_3,\mathbf{r}_1)}{\rho(\mathbf{r}_1)\rho(\mathbf{r}_2)\rho(\mathbf{r}_3)}.$$

Similarly Wick's theorem extends to higher-orders, but we have omitted them for brevity.

Observing ideal bunching amplitudes requires the correlation length at the detector to be significantly larger than the detector resolution [20], which for massive thermal particles has proved challenging with the maximum bunching amplitudes reported in the literature being at most ~20% [$g^{(2)}(0) \approx 1.2$] of the ideal value of $g^{(2)}(0) = 2$ [21] (see Methods in Supplementary Information for the definition of the experimentally measured, volume integrated two-body correlation function at zero interparticle separation). In order to observe ideal bunching amplitudes, we have employed an ultra-cold cloud of partially transversely condensed [2,22] $^4$He* atoms in a strongly confining optical dipole trap (see Fig. 1). This unique cloud can produce ideal bunching amplitudes even with the limited resolution available with current delay-line detectors (~100 microns spatially, ~1 ns temporally). This is due to the gas being transversely coherent, but longitudinally incoherent, where interference between the many longitudinal modes occurs over the entire transverse extent of the cloud.

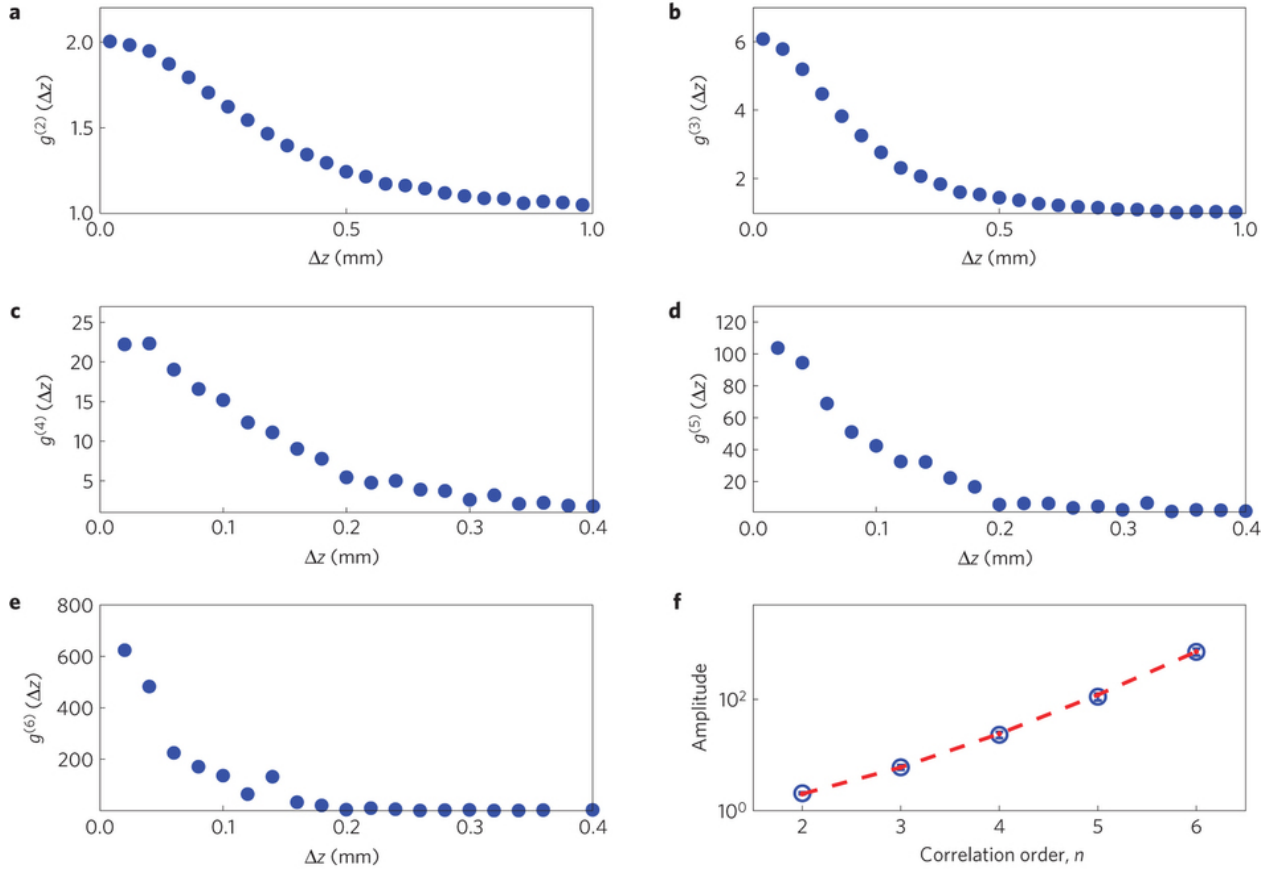

**Figure 2. Many-body correlation functions along the longitudinal direction (temporal dimension on the detector).** The data is averaged over the *x* and *y* transverse directions using ~1cm x 1cm spatial bins. A Gaussian fit to the data yields the following peak correlation amplitudes: (a) $g^{(2)}(0) = 2.05 \pm 0.09$, (b) $g^{(3)}(0) = 6.0 \pm 0.6$, (c) $g^{(4)}(0) = 23 \pm 3$, (d) $g^{(5)}(0) = 111 \pm 16$, and (e) $g^{(6)}(0) = 710 \pm 90$. In (f) we show the *n*! scaling (dashed line) of the peak correlation amplitudes.

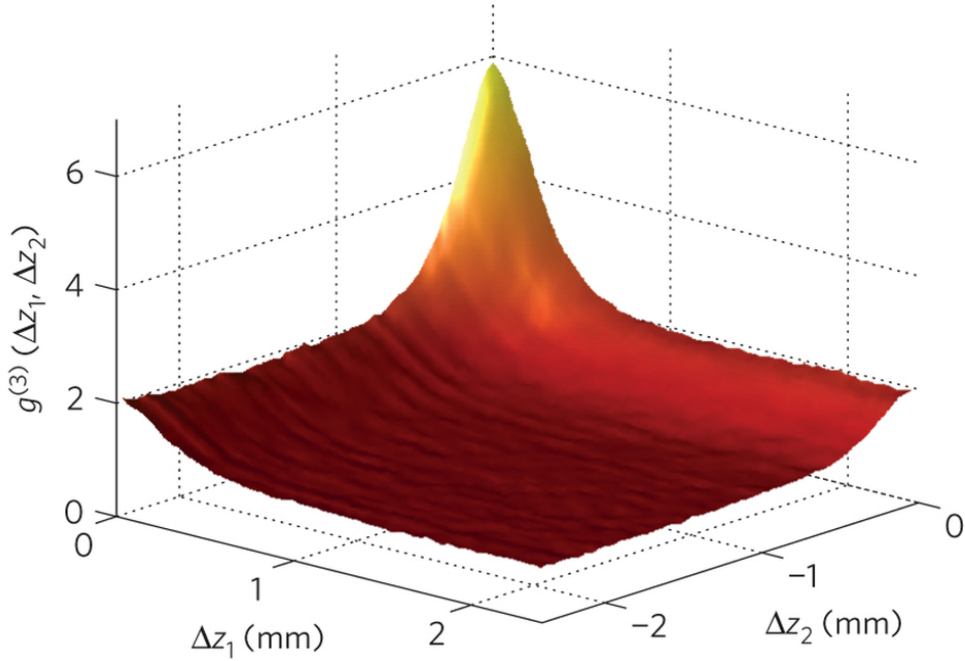

**Figure 3. Surface plot of the three-body correlation function $g^{(3)}(\Delta z_1, \Delta z_2)$.**

The experiment is performed using a novel technique (see Supplementary Information) where we produce ultra-cold clouds of $^4$He* in a highly anisotropic optical dipole trap, aligned with its weak axis in the direction of gravity ($z$-axis). A simplified schematic of the experiment is shown in Fig. 1. The ultracold atomic cloud is at a temperature of just ~60 nK and contains ~330 atoms. To generate the required statistics, over 2000 independent realisations of the experiment were undertaken. We measure spatial correlations after time-of-flight expansion, by dropping the clouds onto a single atom sensitive delay-line detector located ~85 cm below the trap. An average many-body correlation function can then be calculated.

The major result of this paper is illustrated in Fig. 2, where our near ideal many-body correlation functions $g^{(n)}(\Delta z) \equiv g^{(n)}(\Delta z, ..., \Delta z)$ (see Supplementary Information) up to sixth order are displayed as a function of interparticle separation along the longitudinal direction. The peak amplitudes for all the orders are in good agreement with the maximum value of $n!$ expected for the $n$-th order correlation function $g^{(n)}(0)$ for a thermal gas. The peak amplitudes for all the orders are in good agreement with the maximum value of $n!$ expected for the $n$-th order correlation function $g^{(n)}(0)$ for a thermal gas.

To provide further detail on the three-body correlation function $g^{(3)}(\Delta z)$ in Fig. 2(b), we have also plotted in Fig. 3 a full 2D surface plot of the $g^{(3)}(\Delta z_1, \Delta z_2)$-function to which $g^{(3)}(\Delta z) = g^{(3)}(\Delta z, \Delta z)$ is the diagonal cut along $\Delta z_1 = \Delta z_2$. To interpret the surface in Figure 3, one should realise that the three-body interference term is maximum when all three particles are close together (i.e $\Delta z_1 = \Delta z_2 \sim 0$). On the other hand, when one of the particles is taken to large separation (i.e. $\Delta z_i \gtrsim 2$ mm in our case), one recovers the two-body correlation function from this plot along the remaining dimension. The complete fourth-, fifth-, and sixth-order correlation functions, $g^{(4)}(\Delta z_1, \Delta z_2, \Delta z_3)$, $g^{(5)}(\Delta z_1, \Delta z_2, \Delta z_3, \Delta z_4)$, $g^{(6)}(\Delta z_1, \Delta z_2, \Delta z_3, \Delta z_4, \Delta z_5)$ require four-, five- and six-dimensional plots respectively, so for clarity we have plotted these in Figs. 2(c)-(e) similarly to $g^{(3)}$, i.e. for equal spatial separations for all particles.

For atomic systems this is the first reported measurement of the fifth- and sixth-order correlation functions and importantly the measurements of the lower-order correlations are up to two orders of magnitude greater in amplitude than previously reported [13-14, 21]. Remarkably, the measurement of the six-body correlation function demonstrates that the probability of finding six particles at the same location is nearly three orders of magnitude greater than for large separations. Surprisingly, to our knowledge our near ideal sixth-order correlation measurement for massive thermal particles surpasses all measurements for thermal photons.

The peak bunching amplitudes [Fig. 2(f)] of the many-body correlation functions are in agreement with the expected $n!$ dependence of Wick's theorem for thermal particles, thus confirming the validity of quantum theory of boson statistics for massive particles up to the sixth order [1]. Counter-intuitively, the signal to noise ratio is approximately unchanged for all orders. This is due to the signal from the bunching amplitude rapidly increasing which compensates for the increased statistical noise due to a reduction in the number of multi-particle interference events for the higher orders. Ultimately, our finite flux limits the order of the correlation function we can accurately produce. Note, that the two-body correlation function, Fig. 2(a) directly yields the longitudinal correlation length of the gas. However, for correlation functions greater than two-body, Figs. 2(b)–(e), a compression of the width is observed due to our effective bin size increasing.

We emphasise that the nature of the critical transition to a transversely condensed gas, which enabled us to measure higher-order correlations at their ideal limit, is a quantum *degeneracy* driven (rather than interaction driven [23]) transition of an ideal Bose gas confined to a highly anisotropic trap. As discussed in Refs. [2] and [22], this occurs due to the saturation of population in the transversely excited states – hence the term 'transverse condensation'. Our measurements of the transverse properties of the gas, including the fraction of the atoms in the transverse ground state and the transverse correlation functions (to be discussed elsewhere), are indeed in excellent agreement with the predictions of the theory for a harmonically trapped ideal Bose gas. On the other hand, the longitudinal properties of the gas, especially well below the critical temperature, are intermediate between the theory of a highly degenerate ideal Bose gas and a weakly interacting quasicondensate [24]. Given the relatively small number of atoms in our clouds (few hundreds), we expect the finite-size effects to be significant which means that the physics in the longitudinal dimension is dominated by broad crossovers between that of a pure noninteracting gas and of a weakly interacting gas approaching the quasicondensate regime. It is perhaps then not surprising that the measured longitudinal correlation length (394 μm) is significantly larger than that predicted by ideal Bose gas theory (95 μm) but shorter than that expected in the quasicondensate regime (~850 μm) (see Supplementary Information). Nonetheless, we still expect the correlation functions we measure to satisfy the factorial relationship predicted by Wick's theorem, since the equal-point momentum-momentum correlation function is given by $g^{(2)}(\mathbf{k},\mathbf{k}) \approx 2$ for both the ideal Bose gas *and* the weakly interacting quasicondensate regime (see Ref. [24]).

In conclusion, we have measured near ideal many-body correlation functions in a thermal ensemble of ultracold $^4$He$^*$ atoms and demonstrated the validity of Wick's theorem for massive particles to sixth order. Our results show that quantum atom optics experiments can now rival and in some cases exceed the performance of quantum optics experiments. The ability to accurately measure photon-photon correlations in quantum optics has been pivotal in enabling some of the foundational tests of quantum mechanics, such as violations of Bell's inequalities [25-26] and demonstration of the Einstein-Podolsky-Rosen entanglement [27]. Our correlation measurements with ultracold $^4$He$^*$ atoms may pave the way for simular future quantum atom optics tests of the tenants of quantum mechanics for massive particles [12]. In addition, higher-order correlations may play an important role as accurate probes of the 3D condensates [28], as well as intriguing quantum states incorporating lower dimensions [29] and other strongly correlated systems [17]. Finally, they may also provide unambiguous evidence of *p*- and *d*-wave pairings [30], which may offer valuable insights into high-temperature superconductivity.

**Acknowledgements**

A.G.T and K.V.K. acknowledge the support of the Australian Research Council through the Future Fellowship grants FT100100468 and FT100100285.

**Contributions**

S.S.H., R.G.D, and A.G.T. conceived the experiment. A.G.M., S.S.H. and W.R. collected the data presented in this Letter. K.V.K. developed the Bose model and provided theoretical insight into the results. All authors contributed to the conceptual formulation of the physics, the interpretation of the data and writing the manuscript.

**Additional information**

Supplementary information is available in the online version of the paper. Reprints and permissions information is available online at www.nature.com/reprints.

Correspondence and requests for materials should be addressed to A.G.T.

# SUPPLEMENTARY INFORMATION

**METHODS**

The experimental apparatus is similar to that previously described in [1]. In detail, $^4$He* atoms are initially evaporatively cooled in our BiQUIC magnetic trap [2] to just above the BEC transition temperature. A dimple is subsequently formed in the trap by overlapping a red detuned focused laser beam with the magnetic trap and ramping up the laser power adiabatically over a period of 100 ms. The magnetic trap is then switched off leaving a small number of thermal atoms (~$10^4$) in the optical trap at a temperature of ~1 μK. The temperature of the trapped thermal atoms further cooled via evaporative cooling by ramping down the laser power over 200 ms. The anisotropic trap has an aspect ratio of ~100 given by the trapping frequencies $(\omega_x, \omega_y, \omega_z)/2\pi = (2350, 1800, 15)$ Hz.

Correlation functions are measured in the far field by switching off the optical trap, thus allowing the atoms to fall for a time-of-flight $t_{\text{ToF}}$=416 ms onto a detector located ~85 cm below where the arrival time and position of each atom is measured. Single $^4$He* atoms are efficiently detected due to their large internal energy 19.8 eV and long lifetime 7860 s. The detector has a spatial resolution of ~100 microns in the *x-y* plane and an effective spatial resolution in the vertical direction (*z*-axis) of the order of nanometres (temporally a few ns).

*n*-body correlation functions were calculated using the following equation: $g^{(n)}(\Delta z) = \frac{\int d^3 \mathbf{r}\, G^{(n)}(\mathbf{r}, \mathbf{r}+\hat{\mathbf{e}}_z \Delta z, ... \mathbf{r}+\hat{\mathbf{e}}_z (n-1)\Delta z)}{\int d^3 \mathbf{r}\, \langle \rho(\mathbf{r}) \rangle \langle \rho(\mathbf{r}+\hat{\mathbf{e}}_z \Delta z) \rangle ... \langle \rho(\mathbf{r}+\hat{\mathbf{e}}_z (n-1)\Delta z) \rangle}$, where $\hat{\mathbf{e}}_z$ is a unit vector in the *z*-direction, $G^{(n)} = \langle \rho(\mathbf{r}_1)\rho(\mathbf{r}_2)...\rho(\mathbf{r}_n) \rangle$, *n* is the order of the correlation function and $\rho(\mathbf{r})$ is the spatial density distribution on the detector. To produce the second and third order correlation functions shown in the main text, we implement the above equation by counting pairs and triplets respectively within 1 cm x 1 cm bins in the x-y plane and then averaging over the entire ensemble. For orders higher than 3, this procedure is computationally intractable so in such case the data is binned first, using 0.8 cm x 0.8 cm bins in the x-y plane, and then the above equation implemented.

The temperature was derived from a fit to the longitudinal time of flight profile. We use a 3D ideal Bose gas model and fit the curve with a constrained fitting routine. The fitting routine allows the trap frequencies, atom number, and temperature to be varied within bounds that are determined experimentally.

Finally, the longitudinal correlation length for the quasicondensate regime was estimated by relating the spatial correlation length $\Delta z$ after time-of-flight expansion ($t_{\text{ToF}} = 416$ ms) to the characteristic in-trap momentum-momentum correlation scale of $\Delta k \sim 1/2l_\phi$ [3] via $\Delta z = \hbar \Delta k t_{\text{ToF}}/m$, where $l_\phi = \hbar^2 n_{1D}/mk_B T$ is the characteristic phase coherence length, $m$ is the mass of the particle, and $n_{1D}$ is the longitudinal 1D density, which we assume (in the local density approximation) to be the peak 1D density ($n_{1D} \simeq 2.0$ atoms/μm) of our trapped cloud at $T = 63$ nK.